\RequirePackage[2020-02-02]{latexrelease}
\providecommand\DeclareCommandCopy[2]{}
\documentclass[twocolumn,prl,amsmath,amssymb,showpacs,superscriptaddress,floatfix]{revtex4}

\usepackage{graphicx}
\usepackage{bm}
\usepackage{epsf}
\usepackage{xcolor}
\usepackage{natbib}  
\usepackage{amsmath}

\begin{document}

\title{Pattern formation in superconductive vortex systems: two-phase gas-crystal region}

\author{Yu. D. Fomin \footnote{Corresponding author: fomin314@mail.ru}}
\affiliation{Vereshchagin Institute of High Pressure Physics,
Russian Academy of Sciences, Kaluzhskoe shosse, 14, Troitsk,
Moscow, 108840, Russia}

\author{I. D. Irz}
\affiliation{Faculty of Physics, Lomonosov Moscow State University, Leninskie Gory 1/2, Moscow 119991, Russia}

\author{E. N. Tsiok }
\affiliation{Vereshchagin Institute of High Pressure Physics,
Russian Academy of Sciences, Kaluzhskoe shosse, 14, Troitsk,
Moscow, 108840, Russia}

\author{V. N. Ryzhov }
\affiliation{Vereshchagin Institute of High Pressure Physics,
Russian Academy of Sciences, Kaluzhskoe shosse, 14, Troitsk,
Moscow, 108840, Russia}

\date{\today}

\begin{abstract}
In the present paper we critically analyze the results of experimental and computational investigation of inhomogeneous structure of superconductive vortices in 1.5 superconductors. Based on the literature data we propose that the inhomogeneous structure of the vortices can originate from gas-crystal coexistence in the system. We perform additional molecular dynamics simulations of a two-dimensional vortex model and compare the results with the well-known two-dimensional Lennard-Jones system, finding both the equation of state and the sequence of structures to be consistent with gas-condensed phase coexistence.

{\bf Keywords}: molecular dynamics simulation, two-phase region, superconductive vortices, gas-crystal coexistence, phase separation, Lennard-Jones system, 1.5-type superconductors
\end{abstract}

\pacs{61.20.Gy, 61.20.Ne, 64.60.Kw}

\maketitle

\section{Introduction}

It is well known that different systems can form complex patterns, i.e., their structure does not look like a crystal or a liquid but instead forms domains of complex shape (see, for instance, Ref. \cite{domains}). Such patterns have been observed in a wide range of systems of completely different nature: superconducting films \cite{sc-1,sc-2,sc-3,sc-4,sc-6,sc-7,sc-8,zhao}, magnetic systems \cite{mag1,mag2,mag3,mag4}, phospholipids, Langmuir monolayers, and other systems. This means that pattern formation is a general phenomenon rather than a peculiarity of a particular system. At the same time, the mechanisms that lead to pattern formation in systems of such different nature remain poorly understood. More specifically, we describe some examples of pattern formation below. We do not intend to give a complete review of the field, since it is too vast, but rather to give some examples of patterns that form in different systems.

A particular example of pattern-forming systems is provided by 1.5type superconductors (SCs). This type of superconductor has several coherence lengths, which leads to the formation of inhomogeneous structures of Abrikosov vortices in thin superconductive films. Such inhomogeneous structures have been observed in a number of experiments. For instance, in Ref. \cite{m1} magnetic decoration images of vortex structures in MgB$_2$ and NbSe$_2$ were obtained (see Fig. 1(a) and (b)). The authors observed a relatively homogeneous density distribution in the case of NbSe$_2$, while the vortices in MgB$_2$ demonstrated clustering, i.e., denser and sparser regions. These experimental results were complemented by molecular dynamics simulations based on a twogap Ginzburg-Landau (GL) theory (Fig. 2 of Ref. \cite{m1}), which gave similar results. Fig. 3 of Ref. \cite{m1} demonstrates the formation of a stripe phase in MgB$_2$. Similar stripes are also found in the simulations (panel (c) of the same figure).

The authors continued GL simulations of 1.5-SCs in Ref. \cite{m2}, where they found that the vortices can form different patterns — clusters of various sizes and stripes (see Fig. 2 of Ref. \cite{m2}) — and compared them with experimental data. Scanning Hall probe microscopy images of vortex structures in MgB$_2$ and NbSe$_2$ are presented in Ref. \cite{m3} (see Figs.1 and 3–5). They confirm the presence of density inhomogeneities, namely the formation of clusters and stripes in the system.

Inhomogeneous distribution of vortices was also found in other superconductors: BaFe$_{2-x}$NixAs$_2$ (x=0.1 and 0.16) by bitter decoration
method \cite{m4}, Sr$_2$RuO$_4$ and PrOs$_4$Sb$_{12}$ \cite{m5} by SQUID microscopy.

An inhomogeneous distribution of vortices was also found in other superconductors: BaFe$_{2-x}$NixAs$_2$ (x=0.1 and 0.16)) by the Bitter decoration method \cite{m4}, and Sr$_2$RuO$_4$ and PrOs$_4$Sb$_{12}$ \cite{m5} by SQUID microscopy.


As can be seen from the discussion above, there is several pieces of experimental evidence for an inhomogeneous structure of vortices in 1.5-type superconductors. These results are supported by a number of molecular dynamics simulation studies. It has been argued in a number of papers that the interaction of vortices should be repulsive at small separations and attractive at large ones (see, for instance, Ref. \cite{brandt}). A particular form of the interaction potential proposed in several publications consists of the difference of two modified Bessel functions \cite{brandt,kramer}. A particularly careful study of such a system has been undertaken in Ref. \cite{zhao}, which we discuss below. Importantly, although the experimental system is three-dimensional, the vortices can be efficiently considered as two-dimensional (2D) objects. Therefore, simulation studies deal with 2D systems.

A 2D system of particles interacting via the potential:

\begin{equation} 
V_{ij}= V_0 \left( \frac{a}{b} K_0(br_{ij}/\lambda) - K_0(r_{ij}/\lambda) \right),
\end{equation}
where $r_{ij}$ is the distance between two particles, $K_0$ is the zero-order modified Bessel function,
and $V_0$ and $\lambda$ set the energy and length scales, respectively, is considered in Ref. \cite{zhao}. Such potential was developed from
the Ginzburg-Landau theory to describe the interaction between Abrikosov vortices in a superconductor \cite{kramer}.


A search for the ground state structure of the system of vortices interacting via potential (1) 
was performed in Ref. \cite{zhao}. For certain values of $a$ and $b$, the potential develops a minimum at a distance $r_c$. In Ref. \cite{zhao}, certain values of $b$ and $r_c$ were chosen and $a$ was calculated as $a=\frac{K_1(r_c)}{K_1(br_c)}$. Then, the patterns in the system were studied at different $r_c$ and fixed $b$. To this end, molecular dynamics simulations were performed with a Langevin thermostat while gradually reducing the temperature. Simulations at fixed density and varying $r_c$, and at fixed $r_c$ and varying density, made it possible to reveal several types of patterns in the system: small nearly circular clusters, large circular clusters (referred to simply as clusters in Ref. \cite{zhao}), a network with voids (labyrinths in Ref. \cite{zhao}), long stripes of the condensed phase separated by vacuum (stripes in Ref. \cite{zhao}), and a deformed triangular lattice, e.g., with diluted regions (deformed triangular lattice in Ref. \cite{zhao}). All these structures can be seen in Figs. 2-4 and 9-10 of Ref. \cite{zhao}. Figure 5 of Ref. \cite{zhao} shows the appearance of these structures in the $r_c$-density plane. An attempt to find a criterion for the formation of different structures based on the radial distribution functions (RDFs) of the system in these regimes was made in Ref. \cite{zhao}.

Similar results were reported for other models of superconducting films. For instance, the phases of clusters (called 'clumps'), stripes, and anti-clumps (a triangular lattice with voids) were found in a system with an interaction potential of the form
$U(r)=A/r-B \cdot e^{-\kappa r}$ \cite{bishop}, yielding results qualitatively similar to those discussed above.

It can be seen from the discussion above that molecular dynamics simulations of vortices in a superconductor support the formation of inhomogeneous phases in the system. At the same time, to the best of our knowledge, most papers relate the formation of such inhomogeneous structures to the complex shape of the interaction potential. However, it is common knowledge that many simple systems do form inhomogeneous structures. The most familiar example is the two-phase region of the liquid–gas transition, i.e., two phases coexist in the same macroscopic system: drops of liquid surrounded by regions of gas. A less familiar example is a supercritical fluid, i.e., a system at a temperature above the critical point of the gas–liquid transition. Depending on the density, it can be similar either to a gas with liquid-like clusters (low density) or to a liquid with voids (high density).

In the present paper, we perform molecular dynamics simulations of several 2D systems to check whether the inhomogeneities in models of superconductors can be associated with a two-phase region in the system.

\section{System and Methods}

Two systems are simulated in the present work. First, we study a well-understood reference system. This step allows us to establish the behavior expected in a simple system. Then we simulate a model of a superconducting (SC) film and compare the results with those for the simple system.

As an example of a simple case, we simulate a 2D Lennard-Jones (LJ) system:

\begin{equation}
U(r)=\varepsilon \left( \left( \frac{\sigma}{r} \right)^{12} - \left( \frac{\sigma}{r} \right)^{6} \right).
\end{equation}
The parameters $\varepsilon$ and $\sigma$ are used as the units of energy and length, respectively, and all quantities for the LJ system
are expressed in reduced units. The cut-off radius of the LJ system was set to $r_c=2.5$. No tail corrections were employed.
The phase diagram of the 2D LJ system is well known. It contains the crystal, hexatic, liquid, and gas phases. The critical point of the gas-liquid transition is located at $T_c=0.49$, $P_c=0.02858$, and $\rho_c=0.36$ \cite{widom-2d}.

As a model of superconductor, we employ a system with the potential of Eq. (1) 
In reduced units, this potential can be written as \cite{zhao}:

\begin{equation} 
V_{ij} = \frac{V_{ij}}{V_0} = \frac{a}{b} K_0(br_{ij}') - K_0(r_{ij}').
\end{equation}

We use potential (3) 
with $a=1.33425$ and $b=1.1$, which corresponds
to $r_c=2.38$ in the notation of Ref. \cite{zhao}. According to Ref. \cite{zhao}, this system should demonstrate inhomogeneous behavior.

In both cases, we simulate the systems of 20000 particles in a rectangular box with periodic boundary conditions. The simulations were performed
in canonical ensemble, i.e., at a constant number of particles N, area of the system (2D volume) V and temperature T. The density and temperature
are chosen in such a way that crystal-gas two-phase region is explored. The details of the simulation methodology are given in Supplementary
materials \cite{supp}.

All simulations described above were performed using the LAMMPS package \cite{lammps}. The visualization of the results was performed using the OVITO software \cite{ovito}.

\section{Results and Discussion}


\subsection{2D Lennard-Jones system}

We begin with the equation of state of the 2D LJ system shown in Fig. (1) 
One can see that the system exhibits a loop. Moreover, the pressure becomes negative for densities below $\rho=1.0$. Such behavior typically appears when a first-order phase transition between a gaseous and a condensed (liquid or crystal) phase takes place. Since the temperature is below the triple point temperature, a transition between the gas and crystal phases is expected in the system.

\begin{figure}
\includegraphics[width=8cm, height=6cm]{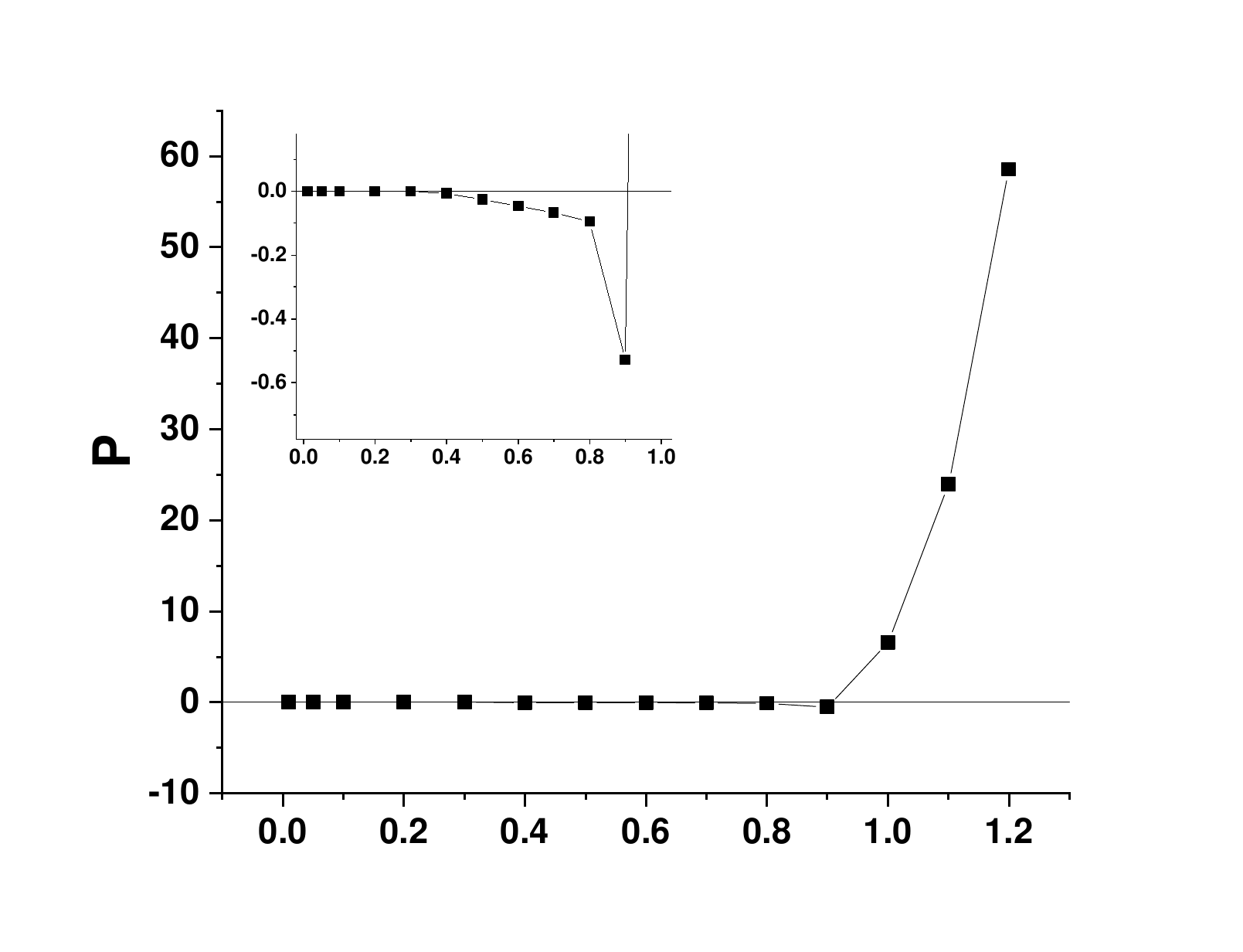}
\caption{Equation of state of 2D LJ system at $T=0.1$. The horizontal line shows the $P=0$ level. The inset enlarges the region of small pressures.}
\end{figure}

Snapshots of the system at different densities are presented in Fig. (2) 
Even at the lowest density, $\rho=0.01$, small clusters form in the system. The core of the clusters appears to have a triangular structure. As the density increases, the clusters become larger, and their triangular structure becomes even more evident. Starting from
$\rho=0.7$, the system looks like a condensed phase with voids. The total area of the voids decreases as the density increases. Finally, at
$\rho=1.0$, an almost perfect triangular crystal is formed.

\begin{figure*}
\includegraphics[width=16cm, height=10cm]{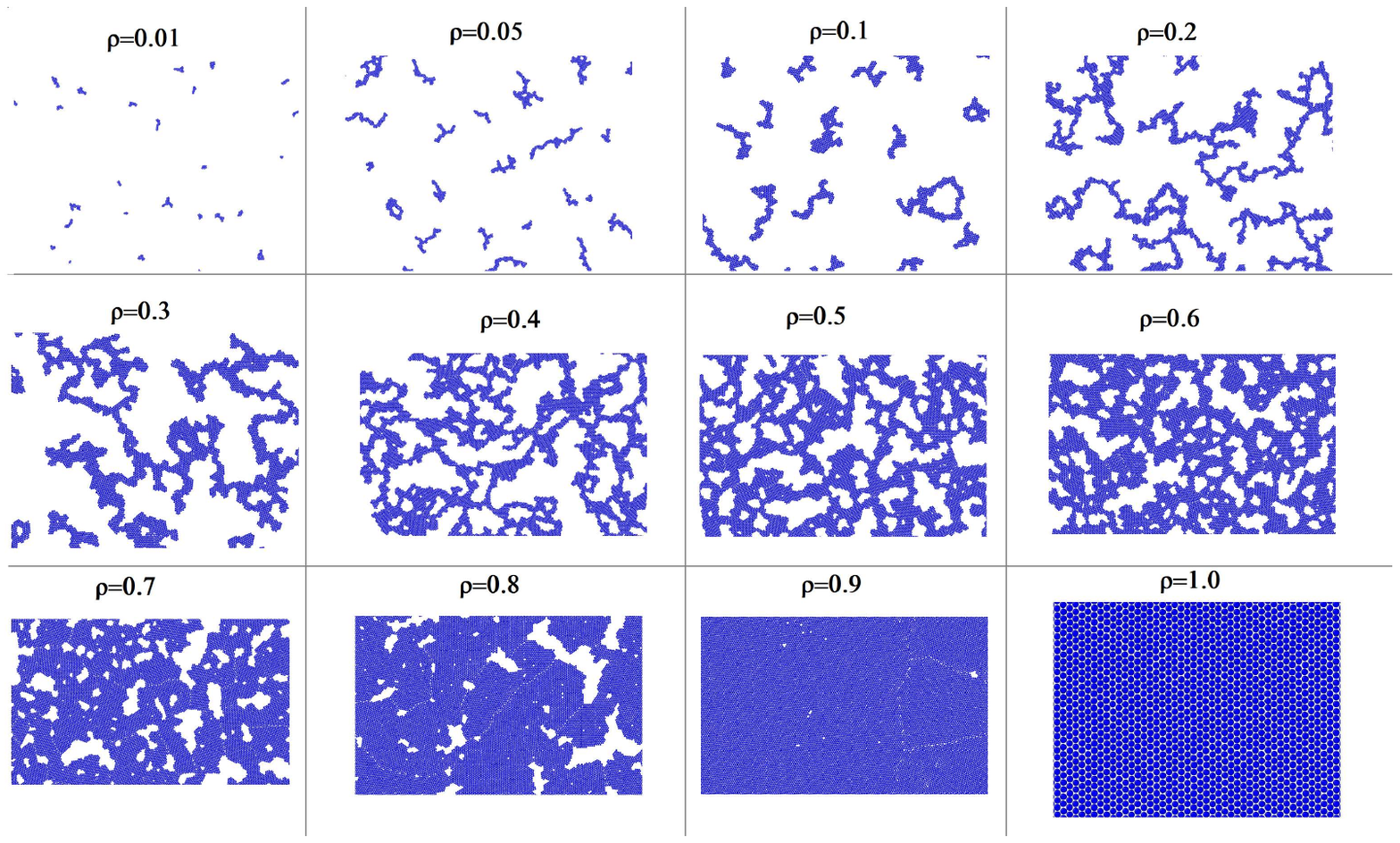}
\caption{Snapshots of the 2D LJ system at $T=0.1$ and a set of densities. Only a part of each system is shown for the sake of clarity.}
\end{figure*}

This sequence of structures is in perfect agreement with the phase diagram of the 2D LJ system and clearly demonstrates the presence of the gas–crystal two-phase region in the system. At the same time, the observed structures closely resemble the snapshots of simulations of vortex structures in 2D SC films. For instance, Fig. 3(c) of Ref. \cite{m1} looks very similar to those for the 2D LJ system at intermediate densities ($\rho=0.2$ - $0.4$). The sequence of snapshots in Fig. 2 of Ref. \cite{m2} corresponds perfectly to that for the 2D LJ system at low densities: small clusters, larger clusters, and stripes.

Based on these results, we see a strong analogy between the inhomogeneous structure of the vortices in 2D SC films and the gas–crystal two-phase region in a simple system.

\subsection{Model of vortices in a superconductor}

At the next step, we simulate a 2D SC film in which the vortices interact via potential (3). 
Figure (3) 
shows the equation of state (EoS) of this system at $T=0.001$. As in the case of the 2D LJ system, a loop with negative pressure is observed. Therefore, a first-order phase transition with a gas phase takes place in the system.

\begin{figure}
\includegraphics[width=8cm, height=6cm]{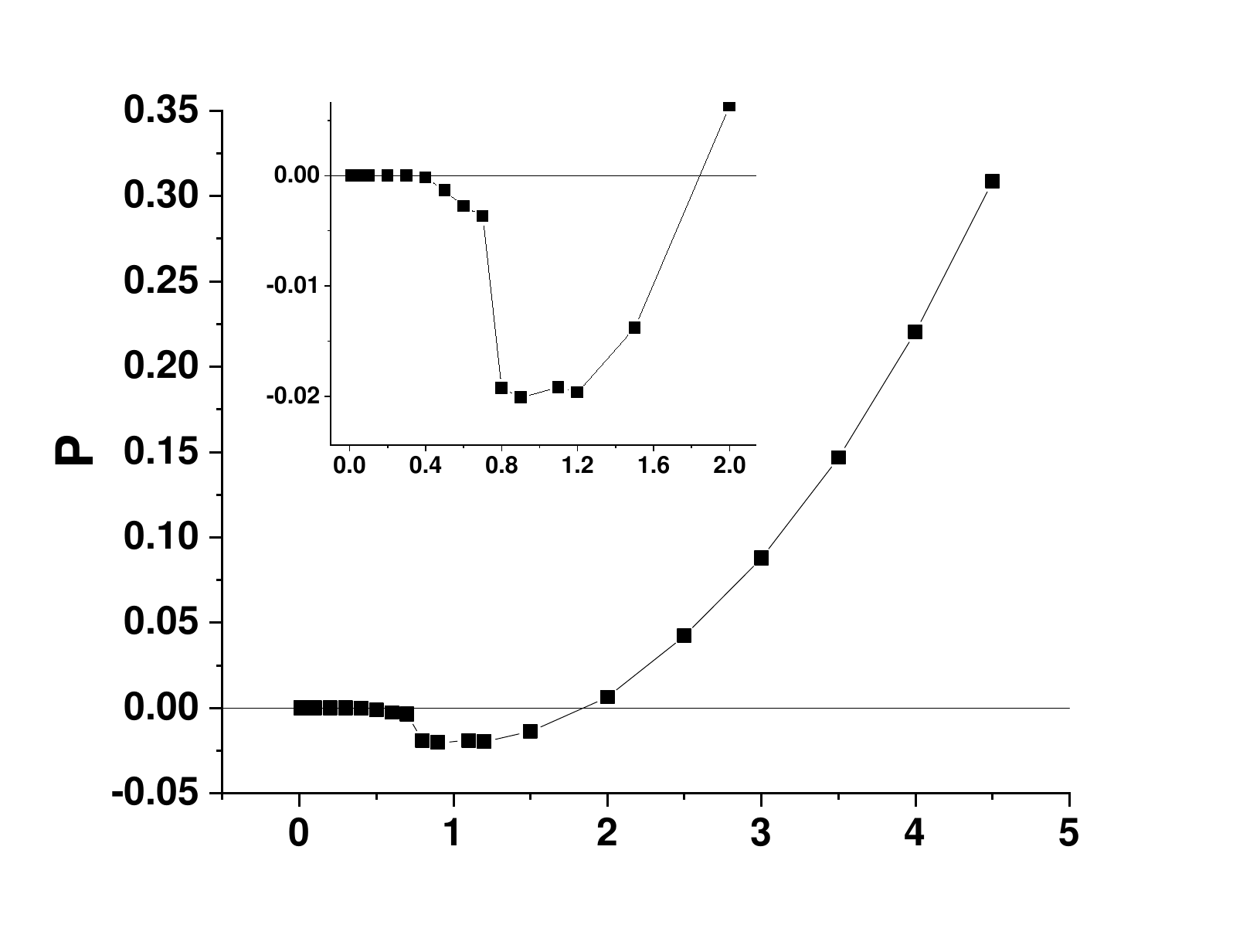}
\caption{Equation of state of the model of superconductor at $T=0.001$. The horizontal line shows the $P=0$ level. The inset enlarges the region of small pressures.}
\end{figure}

Figure (4) shows snapshots of the same system at a set of densities. The patterns are in good agreement with those of Ref. \cite{zhao}. Importantly, Ref. \cite{zhao} mainly investigated the influence of the potential parameters, whereas in the present work we study the structure of the system as a function of density along an isotherm, which allows us to compare the results with those for the 2D LJ system.

As can be seen from Fig. (4), 
circular clusters form for densities up to $\rho=0.5$. The size of the clusters increases with density. Note that the cluster at the bottom of the panel for $\rho=0.5$ is not circular. However, it looks like two circles connected by a neck, which means that the system is still evolving. Although one could simulate the system longer to obtain a better-equilibrated structure, this is not essential in the present case: simulations of two-phase regions suffer from both finite-size and finite-simulation-time problems (see the discussion below). Therefore, it is of principal importance to establish that this is a two-phase region, rather than to determine the exact details of the structure.

At higher densities ($\rho=0.6$ - $0.9$), stripes of the condensed phase surrounded by vacuum are observed. Finally, the system becomes condensed ($\rho=1.1$ and $2.0$).

\begin{figure*}
\includegraphics[width=16cm, height=10cm]{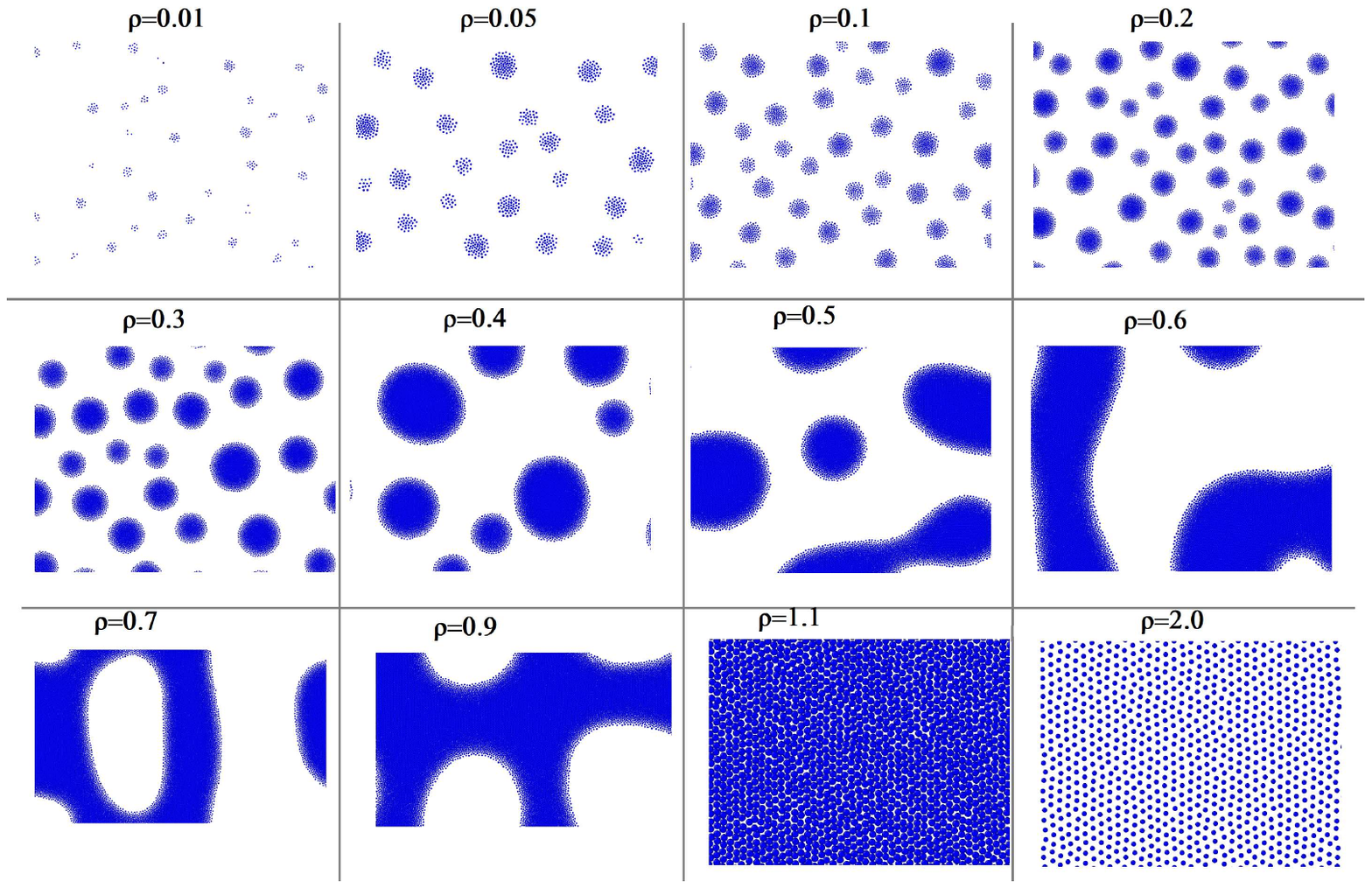}
\caption{Snapshots of the model of the superconductor at $T=0.001$ and a set of densities. Only a part of each system is shown for the sake of clarity.}
\end{figure*}

It is well known that simulations of two-phase regions suffer from finite-size effects: the equation of state may exhibit jumps, and the system forms clusters of different shapes \cite{prestipino-fs}. In the case of a three-dimensional system, the sequence of clusters is as follows: spherical clusters, cylinders, slabs, and condensed liquid with voids. In our recent paper, we analyzed the dependence of these finite-size effects on the size of the system. It was demonstrated that the shape of the clusters themselves can depend on the system size: a cylindrical cluster was observed in a system of 4000 LJ particles at $T=1.0$ and $\rho=0.3$, whereas in the same system with $1 \cdot 10^6$ particles the clusters were elliptical \cite{voids}. Figures (3) and (4) 
demonstrate both effects: a pressure jump when the density goes from $0.7$ to $0.8$, and a change in the shape of the clusters. Therefore, it appears evident that a two-phase region is observed in this case.

The dynamics of a system in a two-phase region can be very slow, which makes it difficult to fully equilibrate the system. A usual equilibration criterion employed in molecular dynamics simulations states that the thermodynamic properties of the system, such as pressure and internal energy, should not change with time. However, in the case of very slow dynamics, the changes can be so slow that they become unobservable within the simulation period. In order to follow the dynamics of the system, we simulated a smaller system of only $5000$ particles at $\rho=0.05$ and $T=0.001$, where circular clusters are observed (see Fig. (4)). 
$r_c=2.0$, which corresponds to the first minimum of the radial distribution function calculated for the system of $20000$ particles at the same density and temperature. Two videos of this long run are provided in the Supplementary Materials \cite{supp}. The first video shows the initial stage of the run. The system starts from randomly distributed vortices, which rapidly merge into circular clusters. The second video shows the whole run, but the frames are taken very rarely (one frame per $1 \cdot 10^6$ steps); otherwise, the video file becomes too large. It can be seen from this video that, after all particles have merged into clusters, the clusters begin to merge as well: when two clusters approach each other, they merge.

The evolution of the system is shown in Fig. (5) 
Panel (a) shows the number of clusters in the system. Initially, the particles do not clusterize, i.e., each cluster consists of one particle. However, the number of clusters rapidly decreases. When plotted on a linear scale, it appears that a saturated value has been reached. This is, however, not true when a logarithmic scale is considered. It is clearly seen that the number of clusters still decreases, but much more slowly.

The same conclusions can be drawn from the size of the largest cluster shown in Fig. 5(b) 
Initially, the size of the largest cluster rapidly increases. Then it reaches certain values that persist for a long time, but suddenly grows due to the merging of two clusters. From this plot, it can be seen that the system continues to evolve. One may expect that all vortices merge into the same cluster in the limit of infinite time. We believe that further simulation of the system would not provide any principal new information. For this reason, we did not continue the simulation.

The results described above have a simple physical explanation. Initially, the particles are randomly distributed, which facilitates the formation of a cluster. The clusters move like Brownian particles. Their diffusion coefficient should be inversely proportional to the square root of the  number of particles in the cluster. Therefore, as the clusters become larger, their motion becomes slower. At the same time, the distance between different clusters becomes larger. These two effects lead to a slowing down of the growth of the clusters, making it extremely slow. In this respect, we would like to note that non-circular clusters have been reported in the literature (see, for instance, Fig. 7(a) of Ref. \cite{zhao}). From our point of view, these clusters represent an intermediate state of the merging of two clusters, i.e., the system is still evolving, and such a structure cannot be considered as the equilibrium one.


\begin{figure}

\includegraphics[width=6cm, height=4cm]{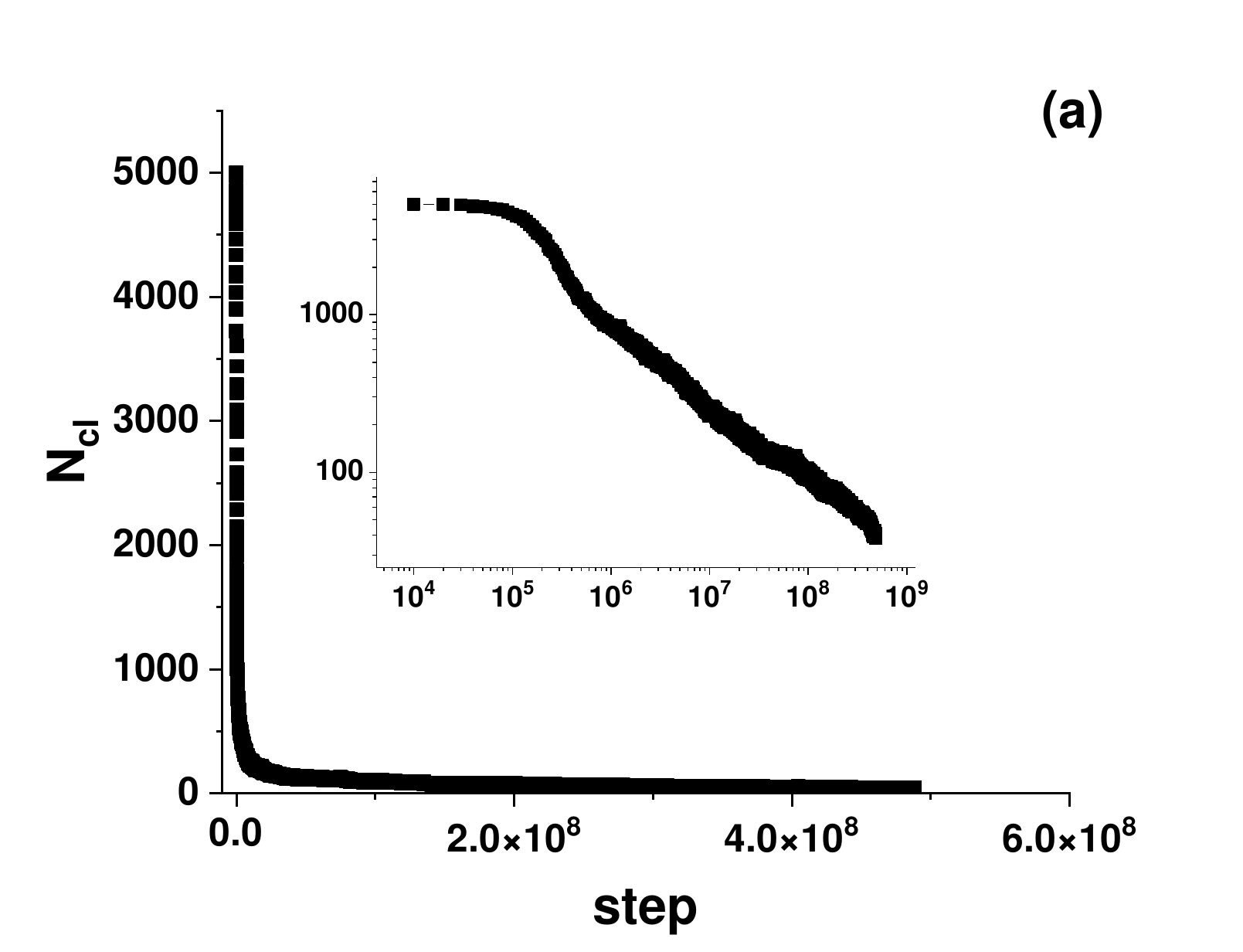}

\includegraphics[width=6cm, height=4cm]{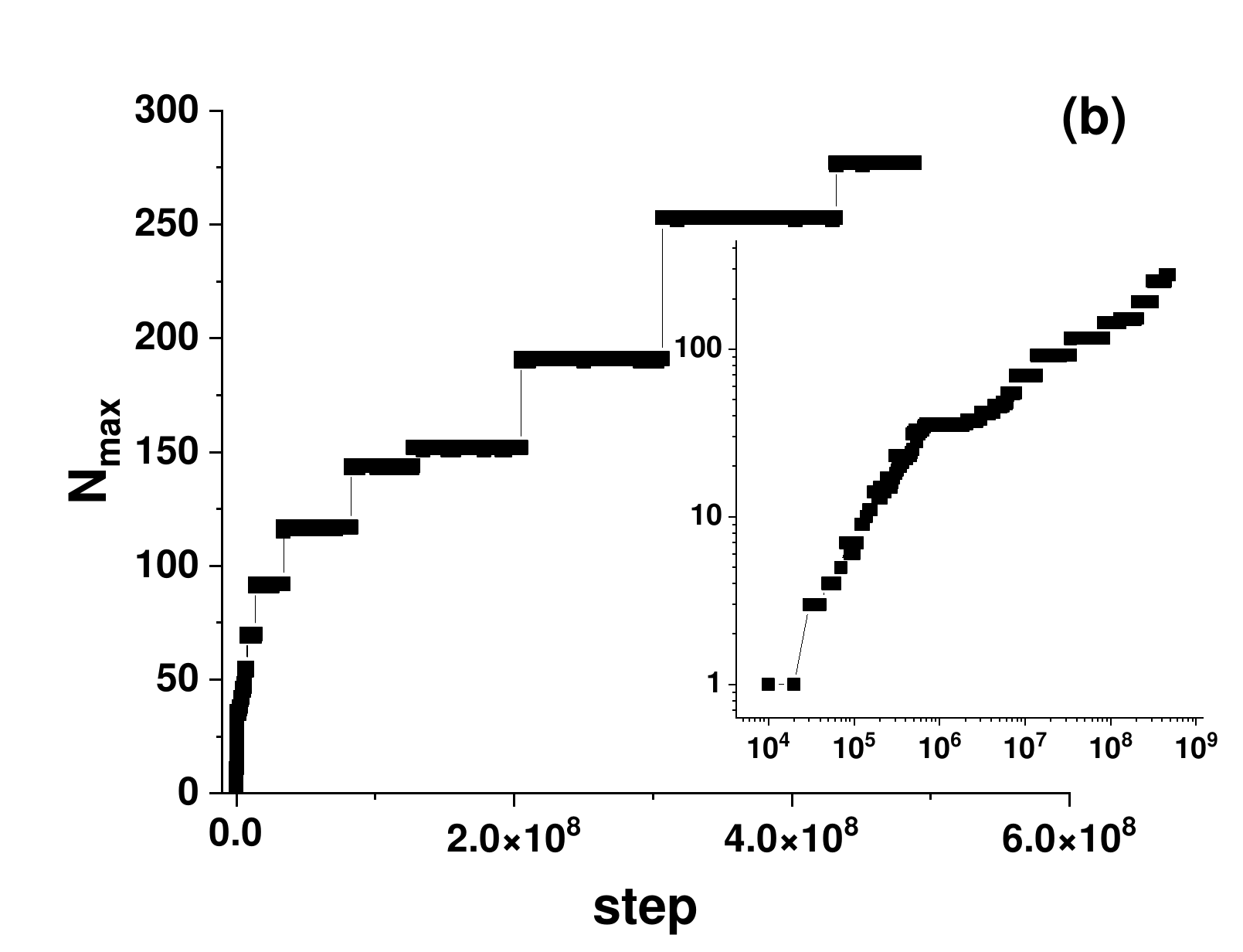}

\caption{(a) Number of clusters and (b) the size of the largest cluster in the long run at $\rho=0.05$ and $T=0.0001$. The insets
show the same data in log-log scale.}
\end{figure}

\bigskip

Based on the discussion above, we believe that the two-phase nature of the inhomogeneous structure of vortices in the models of 1.5-type SC films has been demonstrated. Having established this fact, let us return to the experimental paper \cite{m3}, where the vortex structures were observed by scanning Hall probe microscopy. Figure 2 of this paper reports the nearest-neighbor distance between the vortices in two materials: NeSe$_2$ and MgB$_2$. The local induction $B$ is defined as the number of vortices per total area of the film, i.e., it corresponds to the vortex density. The intergroup nearest-neighbor (NN) distance (between groups) and the intragroup NN distance (within the same group) were measured, and it was found that the intragroup NN distance demonstrates a strong dependence on the magnetic field, whereas the dependence of the intergroup NN distance on the magnetic field is very weak. This is, indeed, the behavior perfectly consistent with the picture of a two-phase region: when the total density of the system becomes larger, the densities of both components (gas and liquid, or gas and crystal) remain constant, but the size of the clusters of the condensed phase increases, which leads to a decrease in the distance between them. Therefore, our interpretation of the inhomogeneous structure of superconductive films appears to be consistent with the experimental data of Ref. \cite{m3}.



\subsection{Limitations and outlook}

It should be emphasized that our goal is to establish the mechanism of inhomogeneous structure formation (gas-condensed phase separation), not to quantitatively reproduce the phase diagram of a specific material. The potential parameters in Eq. (3) are taken from Ref. \cite{zhao} and are not tied to $V_0$ and $\lambda$ for MgB$_2$ and NbSe$_2$; therefore the temperature $T=0.001$ in reduced units cannot be converted to kelvin, and we do not claim that the experimental patterns of Refs. \cite{m1,m2,m3} lie exactly inside our two-phase region. We claim something weaker: the topology of the structures (clusters \ensuremath{\rightarrow} stripes \ensuremath{\rightarrow} condensed phase with voids) as density varies qualitatively matches the sequence expected in a two-phase region, and matches the sequence in the well-studied 2D LJ system, where the location of the two-phase region is known independently. Quantitative calibration to a specific superconductor requires a separate study.

The choice $T=0.001$ is motivated by two considerations. First, at such a low temperature the two-phase region is widest and the gas phase is most dilute, which makes the phase separation visually and thermodynamically most distinct. Second, we deliberately work in a regime where the kinetics are slow, which allows us to separate the thermodynamic fact of the two-phase region's existence (the loop in the equation of state, Fig. 3) from the kinetic details of the morphology. We do not claim that $T=0.001$ corresponds to the temperature of the experiments of Refs. \cite{m1,m2,m3}; we use it as a 'deeply two-phase' regime to demonstrate the mechanism. The width of the coexistence region in density at this temperature is shown in Figs. 3–4; its temperature dependence is a subject for further work.

The long run (Fig. 5) shows that the system does not reach equilibrium within the available time: the number of clusters continues to decrease on a logarithmic scale, and the size of the largest cluster grows in jumps. This means that the snapshots in Fig. 4 represent kinetically arrested intermediate states rather than equilibrium configurations. We emphasize that this does not cast doubt on the existence of the two-phase region - that is established by the equation of state (Fig. 3) and by comparison with the 2D LJ system — but it means that the detailed morphology of the clusters (in particular, non-circular shapes such as 'two circles connected by a neck') reflects coalescence dynamics rather than the equilibrium shape. For the same reason, we do not interpret the exact cluster shapes in the experiments of Refs. \cite{m1,m2,m3} as equilibrium; the comparison with them is qualitative.

\section{Conclusions}

The present study considers inhomogeneous structure of vortices in 1.5-type superconductors. We perform
molecular dynamics simulations of a model of a superconductor in the literature to monitor the structure
of the system at different densities and compare the results with the ones for a well-known model - 2D LJ system.
From the equation of state of the system and comparison to the 2D LJ system we conclude that complex patterns in the
model of superconductor are induced by phase separation of the system into a gaseous and a condensed phase (either
liquid, or crystal depending on the temperature). This conclusion is consistent with available experimental data. See the Limitations and outlook section for the scope of these conclusions.

\section{Acknowledgments}

This work has been carried out using computing resources of the federal
collective usage center Complex for Simulation and Data Processing for
Mega-science Facilities at NRC "Kurchatov Institute", http://ckp.nrcki.ru/.
The work was supported by the Russian Science Foundation (Grant 25-22-00876).

\end{document}